# Differences in Emoji Sentiment Perception between Readers and Writers

Jose Berengueres
*College of IT*
*UAE University*
Al Ain, UAE
jose@uaeu.ac.ae

Dani Castro
*IT Department*
*Happyforce S.L*
Barcelona, Spain
dani@myhappyforce.com

**Previous research has traditionally analyzed emoji sentiment from the point of view of the reader of the content not the author. Here, we analyze emoji sentiment from the point of view of the author and present an emoji sentiment benchmark that was built from an employee happiness dataset where emoji happen to be annotated with daily happiness of the author of the comment. The data spans over 3 years, and 4k employees of 56 companies based in Barcelona. We compare sentiment of writers to readers. Results indicate that, there is an 82% agreement in how emoji sentiment is perceived by readers and writers. The disagreement concentrates in negative emoji, where the authors report to feel 26% worse than perceived by readers. Emoji use was not found to be correlated with author moodiness. Authors that use emoji are happier than authors that never use emoji.**

*Keywords—Sentiment; Emoji; Happiness.*

## I. Introduction

Correctly estimating how a person feels from their social media comments is a powerful datum that can enhance applications that range from automated psychological support (woeBot chatbot) to **suicide** and **depression** prevention systems [1, 2]. One (out of the many) ways to estimate the sentiment of a text is to count the number of negative (and positive) words. To calculate this, public benchmarks of word-sentiment pairs are used. Such benchmarks are usually built by asking humans to read and rate tweets or text from -1 to +1 or by assigning values based on linguistic assumptions (AFINN, etc.). Sentiment estimation has proven popular and today such reader-based benchmarks are used widely to estimate sentiment from a reader point of view. Unfortunately, when it comes to author sentiment there are no author-based benchmarks, therefore reader-based benchmarks are used under the assumption that reader sentiment is equivalent to author sentiment. However, how accurate is such assumption? Here, we focus on emoji and introduce an author sentiment benchmark by taking advantage of a dataset that contains comments with emoji that happen to be annotated for happiness of the writer of the comment. We then compare it to the reader-based benchmark built by [3].

*A. Background on emoji sentiment*

A few studies that correlate emoji with sentiment and other variables exist. For example, [4] correlated the emoji people use with the economic development of the country of the tweet. Emoji can also be used to profile the gender of the author of a tweet [5]. Regarding sentiment, a few papers correlate emoji with sentiment by asking humans to rate tweets where a given emoji appears [3,6]. Perhaps, the most comprehensive benchmark is [3]. In [3], the sentiment ($s$) of an emoji is defined as the mean rating of the tweet where the emoji appears. The scale used is $\{-1,+1\}$ where -1 is most negative and +1 is most positive. Finally, different phone makers and operating systems will display the same emoji code differently. For example, Apple's emoji images are different from the original NTTdocomo emoji set, this impacts how emoji are perceived. [7] studied the variations in interpretation depending on the emoji set used (Apple, Android...) In the next section, we explore the dataset answering simple questions such as, is people that use emoji happier than the rest? In the analysis section, we correlate emoji use with two variables: (i) the happiness score at the time of posting, and (ii) the happiness variability (moodiness). Finally, in the discussion section we compare writer sentiment to reader sentiment and highlight the differences.

## II. Data

In a closely related paper by our group, we developed a model that predicts employee turnover from how an employee used a happiness self-reporting mobile application in the context of company feedback. Then, we determined the top risk factors of employees that turnover by taking advantage of the top predicting features. Here, we use the same data source with new samples that have been added since the time of publication of [8]. The comments are completely anonymized (words have been removed, only emoji remain). The csv files and R code are available at https://github.com/orioli/emoji-writer-sentiment

*A. Data source*

Data used here was collected from 2014-05-10 to 2017-03-08. The data was generated by employees that work(ed) at 56 different companies. The companies belong to one of the following sectors: e-payment start-up, IT consulting services, retail, manufacturing, services, tourism or education. About half of the companies are multinational and the other half are Barcelona-based companies. Employee data was collected in the framework of corporate feedback as part of their companies' kaizen initiatives. The bulk of the employees used this mobile application in Spain (Barcelona area). More than 90% are Spanish nationals. The comments are written in various languages: 97% in Spanish, 2% in English, 1% Catalan. The data consisted of four tables: votes, comments, likes, employee churn. Here, we will focus on two of them: votes and comments. An in-depth review of the dataset is available at [8].

## B. User flow

A happiness *vote* was obtained when an employee opened the app and answered the question: *How happy are you at work today?* To vote, the employee indicates their feeling by touching one of four icons that appear on the screen. See 1st screen, Table 1. The UI of the English version is shown in the table. The default UI was in Spanish language. After the employee indicates their happiness level, a second screen appears where they can input a text explanation (usually a suggestion or comment), this is the comments table. Finally, in a third screen employees can see their peers' comments and like or dislike them. Next, we describe and visualize key metrics of each of the tables.

TABLE I.  SCREENS OF THE APP USED TO COLLECT DATA

| Table (Rows) | Feed-back | UI flow |
|---|---|---|
| Happiness votes (398k) | **How happy are you at work today?** <br> - Great: +1 points <br> - Good: 0 points <br> - So-so: -0.5 points <br> - Bad: -1 points | ↓ 1st screen 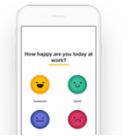 |
| Comments (68,476) | **Comment box** | ↓ 2nd screen 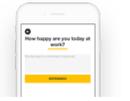 |
| Likes / Dislikes (599,103) | **Anonymous forum** <br> Users can: <br> - **view** comments <br> - **like** a comment <br> - **dislike** a comment | ↓ 3rd screen 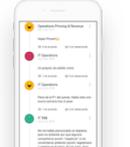 |

## C. Self-reported happiness

Table 2 shows basic statistics about happiness votes aggregated by user. A four-grade scale is used to convert the happiness text categories to numeric. The scale is chosen to map to values as similar as possible to the polarity scale commonly used in sentiment analysis. Therefore, the original happiness labels are converted to numeric as follows: *{bad:-1, so-so: -0.5, good: 0, great: +1}*. The most common answer, good, is assigned zero value. Using this numeric scale, the mean of happiness of all votes is **0.0688**.

Fig. 1 is a histogram of votes recorded. The August and January month dips correspond to the Summer and Christmas vacation period in Spain. Weekly dips are weekends. Fig. 2 is a histogram of votes (includes votes without comments and votes with comments). Company names are anonymized with a one or two-letter code. Fig. 3 is a histogram of app usage derived from the votes timestamps. 0.5 means that the employee used the app once every two days, (days include working days, weekends and holidays). About 10% of the employees used the app every day to report their happiness.

TABLE II.  STATS ON HAPPINESS

| Item | Max | Min | Mean | Median |
|---|---|---|---|---|
| **happiness vote** | **+1** | **-1** | **0.0068** | **0.00** |
| votes per user | 905 | 1 | 57 | 24 |
| mean happiness by user | +1 | -1 | 0.12 | 0.01 |
| standard deviation happiness by user | 1.4 | 0 | 0.41 | 0.43 |

N votes=398k data points

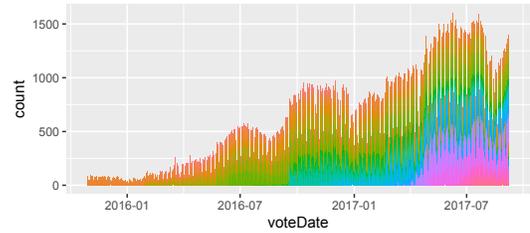

Fig. 1 Count of votes casted daily. Each color is a company. The app is being used by an increasing number of companies.

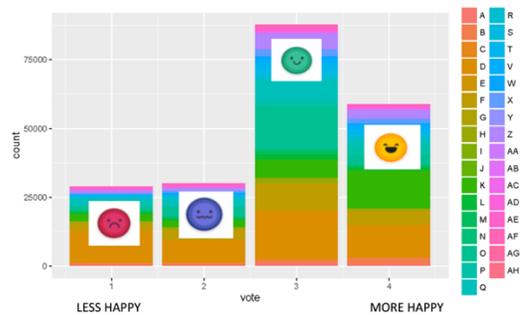

Fig. 2 Count of answers to the question: 'How happy are you at work today?' (1st screen). X axis is index of the categories: *{1:bad, 2:so-so, 3:good, 4:great}*. Period: first 2.5 years.

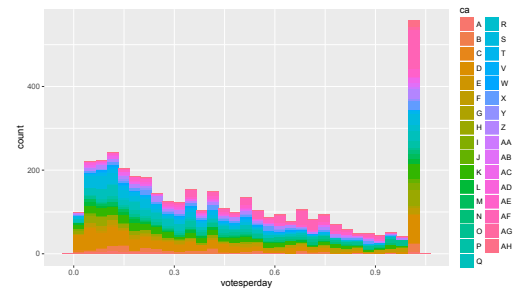

Fig. 3 Usage of the app. About 10% of employees use the app every day.

## D. Comments & Emoji Extraction

Table 3 visualizes the top emoji used by the employees in their comments in the forum. Table 4 visualizes basic statistics of the comments table. In total, 68k comments were recorded with a median length of 58 chars per comment. 5% of comments contain emoji. 962 users used emoji at least once, 4,893 users never used emoji. Out of 63k comments, 3,506 comments contain at least one emoji. 358 different types of emoji are used, (there are 2,666 unique emoji in the Unicode Standard as of June 2017).

TABLE III. TOP EMOJI

| Emoji code | count | % over all emoji | Description |
|---|---|---|---|
| 44f | 884 | 8.8 | Clapping hands |
| 4aa | 879 | 8.7 | Flexed biceps |
| 3fb | 701 | 6.9 | fitzpatrick-1 lighter tone[a] |
| 618 | 471 | 4.7 | Face throwing a kiss |
| 44d | 435 | 4.3 | Thumbs up |
| 389 | 347 | 3.4 | Party popper |

a. a skin tone modifier, not an emoji.

TABLE IV. OVERVIEW OF COMMENTS

| Item | Max | Min | Mean | Median |
|---|---|---|---|---|
| Length of comment in chars | 1k[a] | 1 | 101.7 | 60 |
| Posting date (yy/mm) | 17-09 | 15-02 | 16-12 | 17-02 |
| Count of Emoji per comment that contains emoji | 14[b] | 1 | 2.4 | 2 |
| Author happiness when comment does contain emoji | +1 | -1 | 0.38 | 1 |
| Author happiness when comment does not contain emoji | +1 | -1 | 0.17 | 0 |
| Count of emoji written / user that uses emoji | 268 | 1 | 10 | 4 |
| Emoji per comment of users that use emoji | 49 | 1 | 4.17 | 2 |
| % of comments with emoji by users that used emoji at least once | 90% | 0.4% | 20.1% | 15% |
| Happiness on days when user also posts a comment | +1 | -1 | 0.18 | 0 |

a. comments of longer length are excluded
b. comments with more than 14 emoji are excluded

Fig. 4 visualizes the distribution of count of emoji per comment in the 3,850 comments with emoji that were detected. Comments with more than 15 emoji or longer than 1k char where omitted as outliers. Emoji as html entities where extracted with the following regex expression '&#x([a-zA-Z1-9]{5})'. 10,088 occurrences were detected. As stated before indirectly, of the 63k comments 59k contain no emoji and the rest contain varying amounts of emoji. Fig. 5 compares seven of these groups. Top: shows how the ratio of likes that a comment receives is correlated with the number of emoji in a comment. Bottom: is same looking at the total number of likes received by comment. In both cases, statistically significant differences between groups where not found. In other words, emoji in a comment does **not** influence the sum of likes a comment gets nor the likes to dislikes ratio (see also social approval, likability in [6]).

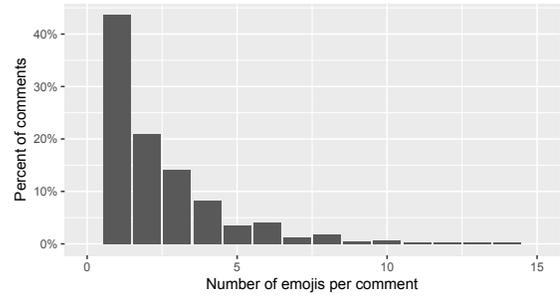

Fig. 4 Of the 3,850 comments that contain emoji, 43% include only one and 57% include multiple emoji.

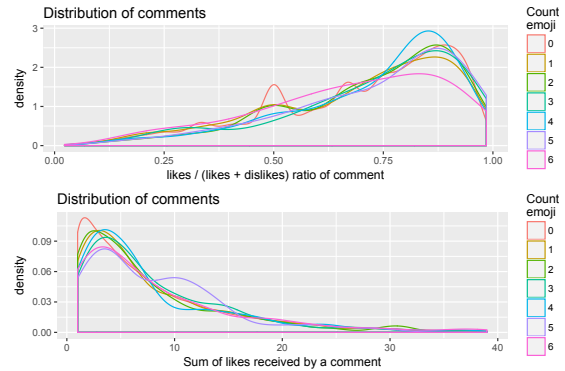

Fig. 5 Top: X is the like ratio. Bottom: X is count of likes received. Y is density of comments. Color label is number of emoji they contain. The number of emoji used in a comment is not correlated with the like ratio or the sum of likes it will receive.

III. ANALYSIS

Regarding author happiness, we look at two parameters: (i) the **mean** happiness and (ii) **moodiness**, the **daily** change in happiness compared to the author's historic mean. We call this change in happiness that occurs daily for every user ***bias***. High variance implies larger sentiment spread, swings (moodiness).

A. Quantization noise

Note that because a 4 point happiness categorical scale is used in the answers {bad, so-so, good, great}, quantization noise will be introduced when converting actual happiness to numbers in a discrete scale. From basic signal processing, we can assume this is a typical analog to digital signal quantization scenario where if we assume that the error introduced is not correlated with the happiness value, and assume that the quantization error introduced follows a uniform noise distribution [9], then for a quantization step of size 1.0, the mean of such noise is zero, the variance = 1/12, and standard deviation = 0.29. The variance for a step size 0.5 is 1/16, and the standard deviation = 0.25. In our scale, we have two steps of size 0.5 and one of size 1. In addition, two of the steps are bounded on one side. Therefore, we can take 1/16 as a lower boundary introduced in our happiness variance measurements. Since the mean is zero, we will not worry about this noise except in mean aggregates with sample sizes < 24. On the other hand, the lower boundary of the variance can be subtracted from the measured variance to find a closer approximation to the true variance of the happiness before quantization noise was introduced, see table 6 footnotes.

*B. Happiness drift*

In addition, for simplicity we assume that the mean happiness of users does not drift or decay for the users (employees). Decay in employee satisfaction has been found to be correlated to employment length and has three components that we will ignore here in order to simplify. The three components are: honeymoon effect in new employees, hangover effect in new employees, and steady decay in happiness of all employees that do not change jobs [10]. This effects have been confirmed to appear in this dataset too (not shown here but available as a kernel in Kaggle [8]).

*C. Comments with emoji vs. comments without emoji*

Fig. 6 shows the evolution of happiness of authors on days that they post a comment. Left chart: The Y is the happiness of the author when they posted the comment. X is the date of the comment. Right chart is a density bar plot.

In Fig. 5 we saw that the fact that a comment contains emoji does not affect the like ratio of a comment (note that likes are given by readers of the comment). However, when an author uses emoji they report happier levels than when not. A Kolmogorov test also supports this hypothesis. However, the margin of confidence of the moving averages CI 95, (shown in grey) overlaps for some periods of time. Sample size is N=3,850 for comments with emoji and N=60k for comments not containing any emoji. Fig. 7 is the like Fig. 6 except that Y is now bias as defined earlier. Dots are not shown in colors for aesthetics. The chart to the right is the corresponding density plot of the comments that contain at least one emoji vs. all the comments with no emoji. A Kolmogorv test does **not** support the hypothesis that the two distributions have different means (p-value = e-16). This is, posting a comment with or without emoji is **not** correlated with a significant increase or decrease in bias (moodiness) on the day of posting. Fig. 8 examines the relationship between the long-term happiness mean of a user and how often they use emoji. Users that use emoji more than 50% of the days are hardly unhappy. Table 5 compares emoji users with non-emoji users.

TABLE V. HAPPINESS OF EMOJI USERS VS. NON-USERS

| Item | Does vote belong to an author that uses emoji? | | |
|---|---|---|---|
| | **no,** author never used emoji | **yes,** sometimes | |
| | | used emoji in a comment today? | |
| | | no | yes |
| **Mean happiness (votes)** | **0.105** | **0.27** | **0.38** |
| standard deviation votes happiness | 0.73 | 0.73 | 0.69 |
| N votes | 36934 | 22286 | 3850 |
| N unique authors (user id) | 3714 | 925 | |

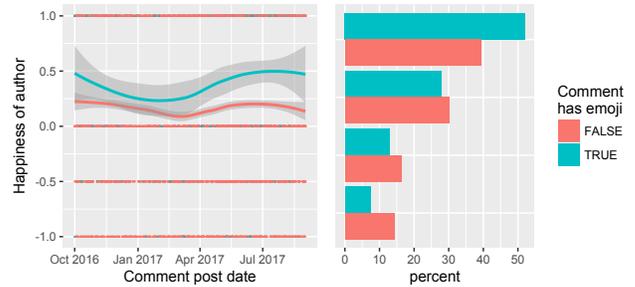

Fig. 6 Evolution of happiness of authors on days that they post a comment. Are authors happier when they use emoji in a comment? Yes, but not all the time. Note that compared to Fig. 2, the most common answer is now {4: great} assigned +1 score.

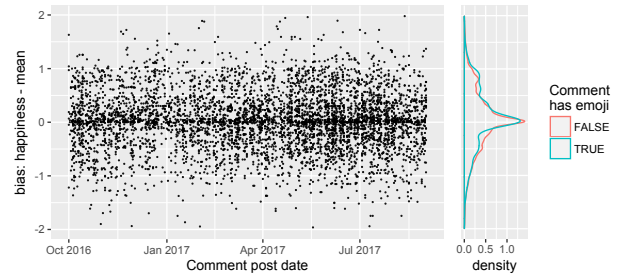

Fig. 7 Y (bias) is the difference in daily happiness minus mean of each author. X is timeline. Labels are same as in fig 6. Red is for comments that contain no emoji. Only 7k points are shown due to computational restrictions.

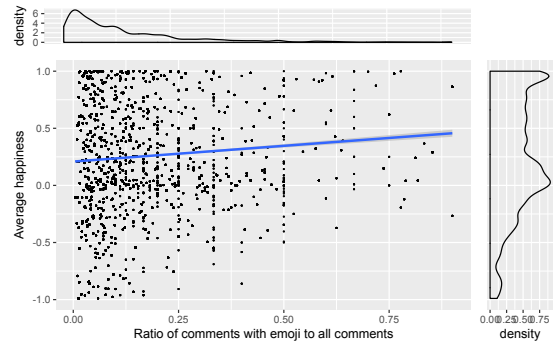

Fig. 8 Each dot represents an emoji user (an author that used emoji at least once), the sample size is N=925. X is the ratio of number of comments with emoji divided by all the comments the author wrote. Y is the mean happiness of the author (using all votes when they commented too). The blue line is a linear regression. R-squared 0.006. Empty quarter: users with a ratio greater than 50% are seldom unhappy.

*D. Estimating sentiment from emoji*

Fig. 9 and 10 show the happiness and bias for 4 groups of comments. Certain specific emoji are powerful predictors of the writer happiness. However, how well can emoji that appear in a comment be used to estimate the sentiment of the author on the day the comment was posted? Fig. 11 answers this question. It is a scatter plot of bias vs. standard deviation of bias. The bias in the y axis is normalized by the standard deviation because it increases readability and to account for the fact that low standard deviation variables are better predictors al else being equal. In this figure, we can see which emoji are good predictors of sentiment change in the author as compared to his usual happiness level, the '*B-day cake emoji*' is associated with a low standard deviation and a positive "boost" in the authors reported happiness on the day it is used. The dancer woman is associated with the largest daily boost followed closely by the '*hugging face*'. Fig. 11 is a scatter plot of mean happiness vs. standard deviation of happiness for the most common emoji in our dataset (counts >50). The plots show that emoji are predominantly positive (Fig. 11). However, the negative emoji have larger bias than positive emoji (Fig. 10).

*E. Reader vs. writer*

In Table 6, for each emoji we list the mean happiness of author (writer sentiment) when they wrote the comment where the emoji appears. Next to it, we list **reader** sentiment provided by [4]. The diff column is the difference. Fig. 13 is a scatter plot of reader sentiment vs. writer sentiment. The grey line is y=x. Points above indicate emoji where writer rating > reader rating. Notice that there are two emoji that are rated negative sentiment by the writers but positive sentiment by readers. These emoji are: crying face and disappointed but relieved face. Another highlight is the pouting face emoji whose writer sentiment is double the reader sentiment, both negative. To estimate the significance of difference in means between reader and writer for a given emoji such as the emoji 'disappointed but relived face', we have that reader sentiment is: 0.122 and no standard deviation (standard deviation) available, sample size N = 341. Writer sentiment is -0.275, standard deviation = 0.634, sample size N = 60. We assume standard deviation reader = standard deviation writer as we don't have the value. Then, applying a Kolmogorov test with simulated samples with the said statistics we have that the difference in means is significant with an alternative hypothesis p-value = 0.0001048.

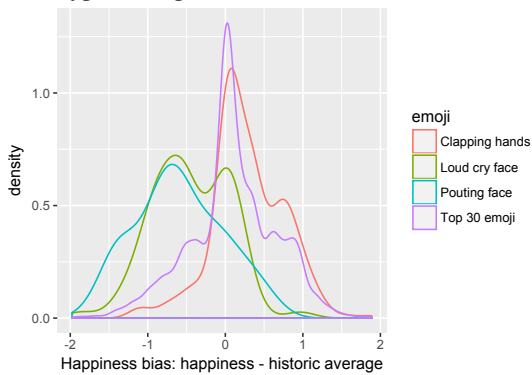

Fig. 9 Distribution of bias. When pouting face emoji is used the authors' happiness level was less than their historic mean 75% of the time.

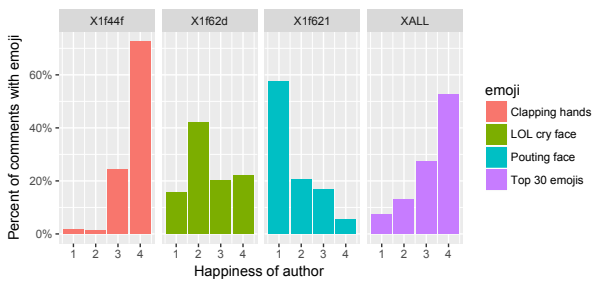

Fig. 10 Bar density plot of 3 given emoji and the top 30 emoji aggregated (purple). When clapping hands emoji is used, the author was at a happiness level of 4 more than 70% of the time. X scale is categorical labels index (1 lest happy, 4 most happy).

*F. Regression*

A linear regression that estimates writer sentiment from reader sentiment (not shown in Fig. 13 but very close to the y=x line shown) yields a multiple R-squared: 0.824. Meaning that 82% of the sentiment variation in the writer can be inferred from reader ratings provided by [3]. However, the intercept is -0.048 meaning that if reader sentiment is used to estimate writer sentiment, we would systematically underestimate happiness by an average of 0.04 points, or normalizing by the standard deviation of writers (0.41), by about 10% of their standard deviation.

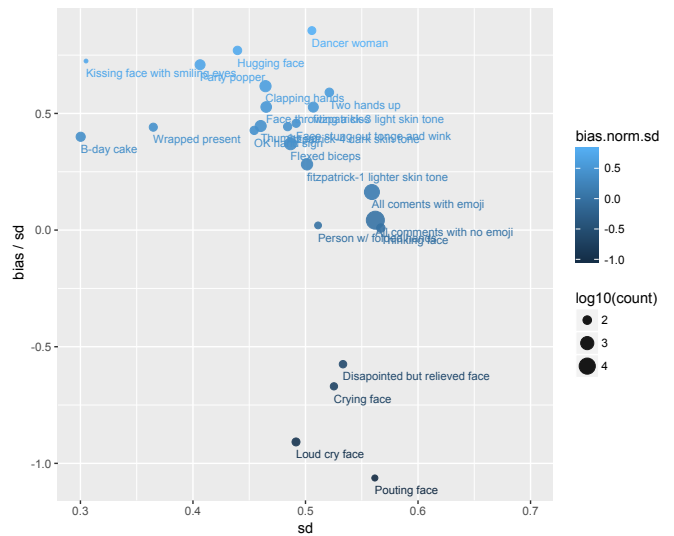

Fig. 11 Scatter plot. X is standard deviation of bias. Y is bias normalized by the standard deviation of bias. Size of dot represents sample size. Writer sentiment shows the greatest swings when the pouting face emoji is used.

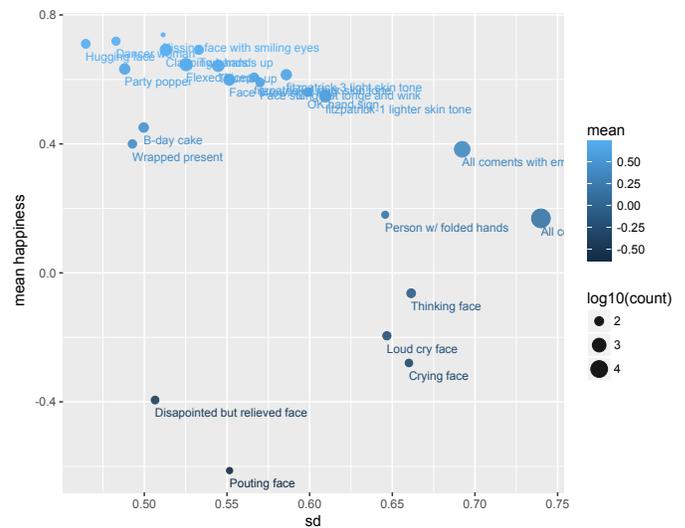

Fig. 12 Scatter plot. Y is mean happiness. X is standard deviation of happiness. Size of dot represents sample size. Comments with emoji are happier and exhibit less variability. Only counts >50 are shown for clarity.

## G. Self-selection of happy users

Fig. 14 shows the happiness density distribution of three groups of votes (i) votes with no comment, (ii) votes with comment (with and without emoji), (iii) votes with comments containing emoji. The groups are ordered by the amount of effort required from the writer. (i) requires the least effort (choosing an icon in screen 1 of the app), and (iii) requires the most effort (voting, typing, choosing an emoji, and perhaps even switching keyboards). Rearranging we can get the more useful Fig.15 where the conditional probability of happiness of author given the case is shown. In Fig.15 we have arranged the cases by growing degree of engagement [11]. From this we see that the probability of happiness {1: great} increases monotonically with engagement. The probability of {4: bad} decreases monotonically with engagement. The probability of {2: good} and {3:so-so} are mixed. However, they would decrease monotonically if they were merged. On the other hand, while Fig. 16 shows that more engagement is correlated with more happiness, this is not always the case in other engagement aspects that can be measured. For example, Fig. 16 shows the relation between comment length and reported happiness on that same say. For long comments (length >20 chars), the longer the comment (engagement), the unhappier (perhaps angrier too) the authors report to feel.

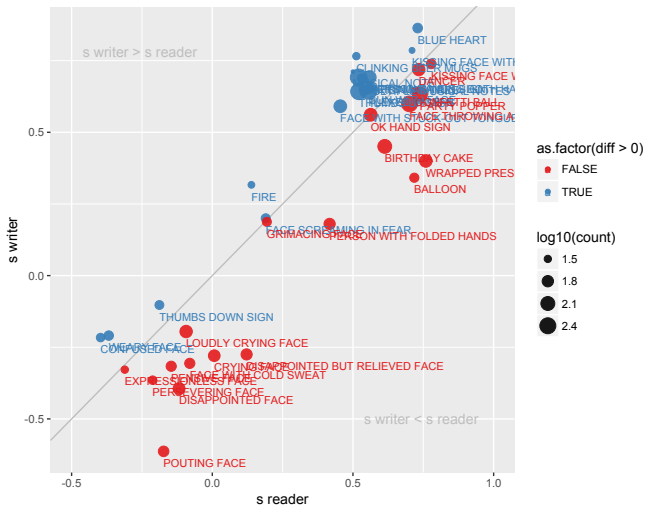

Fig. 13 Y is writer sentiment, X is reader sentiment [4]. Dots are comments having a certain emoji. The grey line is y=x. Red indicates that sentiment perception of reader > writer. Only counts >24 are shown. R-squared = .82

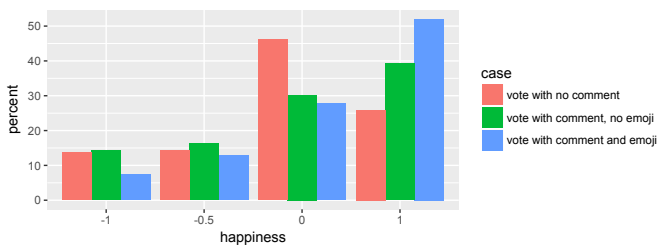

Fig. 14 Self-selection of happy users. The more effort it takes, the less likely to have unhappy users. X is happiness of vote. Y is percentage of votes within each label.

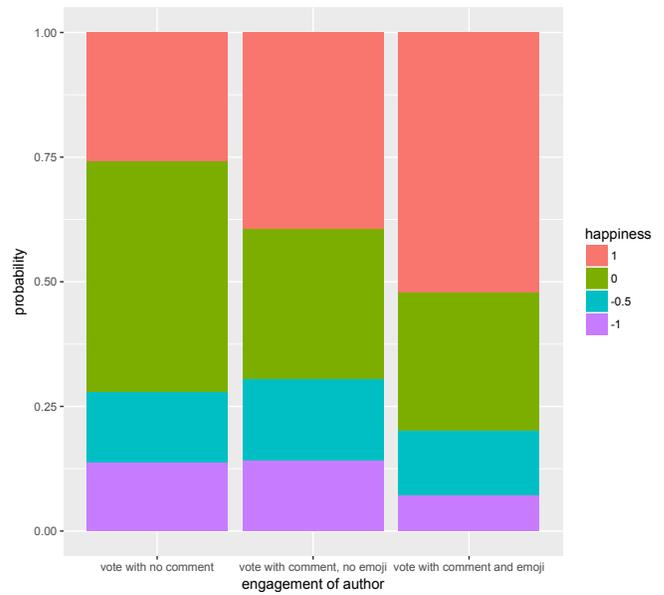

Fig. 15 Conditional probability of happiness given the author engagement. Happiness scale {1: great, 0: good, -.5: so-so,-1: bad}

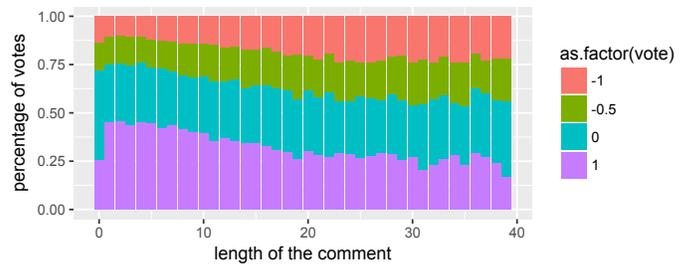

Fig. 16 X is length of comment, units is 10x char

## IV. DISCUSSION

### A. Emoji use is correlated with happier users

Authors that use emoji self-report happier levels than authors that never use emoji. While both groups show similar variability, authors that never use emoji self-report an average happiness of +0.10 while authors that use emoji report +0.27 on days that they did not use emoji in their comments and +0.38 on days they do use at least one emoji in their comment. However, correlation does not imply causality. In Fig. 2 we saw that the most common vote for happiness is {3: good}, (assigned 0 points in the sentiment scale). However, when we look at the same histogram considering **only** votes where the user also commented, the most common vote is {4: great}, see also Fig. 16. A self-selection of happy users seems to be happening. Why? A hypothesis is that typing an emoji requires an extra effort by the writer as changing the keyboard from QWERTY to emoji keyboard is required. Thus, the cause of increased happiness in users that use emoji, (as in users that care enough about their jobs to vote and then input a comment as opposed to just voting without inputting a comment), might not be the emoji but the fact that users that care enough to ameliorate and beautify their comment with emoji are more engaged than those that chose not to do so. The fact that there are more positive emoji than negative ones is also a confounding factor. This results regarding engagement are in line with the findings of [8], where low engagement was found to be a predictor of employee turnover.

### B. Moodiness

As we saw in the density plot of Fig. 7, bias (the difference between daily happiness and an author's average long term happiness) is **not** correlated with emoji use. In other words, taking bias as an indicator of moodiness, comments that contain emoji are not moodier nor have higher moodiness variance than emoji-free comments.

### C. Predictive power

Certain emoji are correlated with high levels of unhappiness. For example, the pouting face emoji is correlated with the lowest happiness category {bad:-1} more than 50% of the time it is used.

### D. Writer vs. reader sentiment

In table 6, we compare reader sentiment [3] with writer sentiment (from this dataset). Only emoji with more than 24 counts and that are also listed by [3] are shown. A linear regression was computed to estimate writer sentiment from the reader sentiment. It yields an R-squared = 0.82, meaning there is a 18% of the variability in writer sentiment that is not explained by looking at reader sentiment of [3], (see column "sentiment reader"). In addition, even though the regression's R-squared score is high, the intercept is -0.047. This means that a reader originated benchmark such as [4] consistently overestimates the sentiment of authors by 0.04 points (10% of the standard deviation). This percentage becomes close 15% if we use the approximation of variance where we subtract the lower boundary of the noise variance introduced in the quantization step. The 'Ok sign' is the emoji where readers and writer's sentiment agree the most. On the other hand, the 'pouting face' is the emoji where readers and writers **disagree** the most. Writers perceive it much more negatively than readers do by an average of 0.44 points. At an aggregate level (see right column table 6), we split emoji in two groups: positive and negative, and compute weighted averages by count. For positive emoji (blue heart, kissing face...) writers are more positive than readers by 0.027 points or a non-significant 3.3% of the standard deviation. However, for negative emoji (bottom of table 6), the writers are more negative than readers by an average of 0.166 points (26% of the standard deviation, N sample size = 620). This means that negative emoji is where most of the disagreement between readers and writers occurs. In other words, if [3] is used to estimate author sentiment, for negative emoji we would be underestimating the negativity of what the author feels by 0.166 points. Normalizing this by the weighted standard deviation of negative emoji, this gap amounts equals to a significant 26%.

### E. Skin modifiers

Finally, the Spanish workers used in some cases skin color modifiers in their emoji. The skin modifiers detected with more than 24 counts were (Fitzpatrick tones 1, 3, and 4). Fig. 13 shows no significant differences between skin color and sentiment.

## V. CONCLUSION

We have introduced an author sentiment benchmark of emoji and we have compared it with a writer sentiment benchmark. We found that readers and writers seem to broadly agree on sentiment of emoji (84%) even though the reader's source of comments were tweets and the writer's source of comments was an anonymous work-happiness monitoring app where comments refer to work suggestions. The largest disagreement between readers and writers occurs for negative emoji (sad face, pouting face, etc...). For this group of emoji, the authors report to feel 26% worse than what readers perceive in terms of standard deviation. Emoji use was not found to be correlated with author moodiness. Emoji use was found to be correlated with more happiness within users and between users. We hope this results will improve the accuracy of sentiment estimation of authors and writers when emoji are used.

TABLE VI. COMPARISON OF EMOJI SENTIMENT IN READERS VS. WRITERS

| | code 1xf | sentiment writer | sentiment reader[a] | standard deviation writer[b] | count[c] | emoji description | s.reader - s.writer | weighted average s.reader - s.writer |
|---|---|---|---|---|---|---|---|---|
|  | 499 | 0.864 | 0.730 | 0.409 | 44 | BLUE HEART | -0.134 | |
|  | 61a | 0.786 | 0.710 | 0.418 | 28 | KISSING FACE WITH CLOSED EYES | -0.076 | |
|  | 37b | 0.766 | 0.512 | 0.523 | 32 | CLINKING BEER MUGS | -0.254 | |
|  | 619 | 0.739 | 0.778 | 0.511 | 44 | KISSING FACE WITH SMILING EYES | 0.039 | |
|  | 483 | 0.719 | **0.734** | 0.483 | 80 | DANCING WOMAN | 0.015 | |
|  | 3b5 | 0.712 | 0.500 | 0.493 | 26 | MUSICAL NOTE | -0.212 | **positive emoji** |
|  | 64c | 0.692 | 0.559 | 0.533 | 99 | PERSON RAISING BOTH HANDS... | -0.133 | |
|  | 44f | 0.692 | 0.520 | 0.513 | 305 | CLAPPING HANDS SIGN | -0.172 | mean = -0.02 |
|  | 384 | 0.689 | 0.531 | 0.491 | 37 | CHRISTMAS TREE | -0.158 | |
|  | 3b6 | 0.684 | 0.537 | 0.512 | 38 | MULTIPLE MUSICAL NOTES | -0.147 | sd=0.49 |
|  | 31e | 0.654 | 0.558 | 0.502 | 39 | SUN WITH FACE | -0.096 | |
|  | 4aa | 0.646 | 0.555 | 0.525 | 445 | FLEXED BICEPS | -0.091 | |
|  | 38a | 0.645 | 0.721 | 0.486 | 31 | CONFETTI BALL | 0.076 | |
|  | 44d | 0.642 | 0.521 | 0.545 | 288 | THUMBS UP SIGN | -0.121 | |
|  | 389 | 0.632 | 0.738 | 0.488 | 170 | PARTY POPPER | 0.106 | |
|  | 618 | 0.599 | 0.701 | 0.551 | 232 | FACE THROWING A KISS | 0.102 | |
|  | 61c | 0.591 | 0.455 | 0.570 | 88 | FACE WITH STUCK-OUT TONGUE... | -0.136 | |
|  | 44c | 0.561 | 0.563 | 0.599 | 90 | OK HAND SIGN | 0.002 | |
|  | 382 | 0.451 | 0.613 | 0.500 | 122 | BIRTHDAY CAKE | 0.162 | |
|  | 381 | 0.400 | 0.759 | 0.493 | 85 | WRAPPED PRESENT | 0.359 | |
|  | 388 | 0.341 | 0.718 | 0.480 | 41 | BALLOON | 0.377 | |
|  | 525 | 0.317 | 0.139 | **0.835** | 30 | FIRE | -0.178 | |
|  | 631 | 0.200 | 0.190 | 0.791 | 40 | FACE SCREAMING IN FEAR | -0.010 | |
|  | 62c | 0.188 | 0.194 | 0.667 | 40 | GRIMACING FACE | 0.007 | |
|  | 64f | 0.180 | 0.417 | 0.646 | 61 | PERSON WITH FOLDED HANDS | 0.237 | |
|  | 44e | -0.103 | -0.188 | 0.829 | 39 | THUMBS DOWN SIGN | -0.085 | |
|  | 62d | -0.195 | -0.093 | 0.647 | 82 | LOUDLY CRYING FACE | 0.102 | |
|  | 629 | -0.209 | -0.368 | 0.526 | 43 | WEARY FACE | -0.159 | |
|  | 615 | -0.216 | -0.397 | 0.534 | 37 | CONFUSED FACE | -0.181 | **negative** |
|  | 625 | -0.275 | 0.122 | 0.634 | 60 | DISAPPOINTED BUT RELIEVED FACE | 0.397 | **emoji** |
|  | 622 | -0.279 | 0.007 | 0.660 | 68 | CRYING FACE | 0.286 | |
|  | 613 | -0.306 | -0.080 | 0.576 | 49 | FACE WITH COLD SWEAT | 0.226 | mean = |
|  | 614 | -0.316 | -0.146 | 0.675 | 49 | PENSIVE FACE | 0.170 | **+0.16** |
|  | 611 | -0.328 | -0.311 | 0.577 | 32 | EXPRESSIONLESS FACE | 0.017 | sd=0.60 |
|  | 623 | -0.365 | -0.212 | 0.573 | 37 | PERSEVERING FACE | 0.153 | |
|  | 61e | -0.394 | -0.118 | 0.506 | 71 | DISAPPOINTED FACE | 0.276 | |
|  | 621 | -0.613 | -0.173 | 0.552 | 53 | POUTING FACE | **0.440** | |

a. data source [3] http://kt.ijs.si/data/Emoji_sentiment_ranking/

b. To obtain a more accurate standard deviation estimate subtract the quantization noise variance lower boundary, sqrt(sd$^2$-1/16)

c. Only counts > 24 shown and emoji present in both sets

## REFERENCES

[1] Fitzpatrick, Kathleen Kara, Alison Darcy, and Molly Vierhile. "Delivering Cognitive Behavior Therapy to Young Adults With Symptoms of Depression and Anxiety Using a Fully Automated Conversational Agent (Woebot): A Randomized Controlled Trial." JMIR Mental Health 4.2 (2017): e19.

[2] Robinson, Jo, et al. "Social media and suicide prevention: a systematic review." Early intervention in psychiatry 10.2 (2016): 103-121.

[3] Novak, Petra Kralj, et al. "Sentiment of emojis." PloS one 10.12 (2015): e0144296

[4] Ljubešic, Nikola, and Darja Fišer. "A global analysis of emoji usage." ACL 2016 82 (2016).

[5] Martinc, Matej, et al. "Pan 2017: Author profiling-gender and language variety prediction." Cappellato et al.[13] (2017)

[6] Kiritchenko, Svetlana, Xiaodan Zhu, and Saif M. Mohammad. "Sentiment analysis of short informal texts." Journal of Artificial Intelligence Research 50 (2014): 723-762.

[7] Miller, Hannah, et al. "Blissfully happy" or "ready to fight": Varying Interpretations of Emoji." Proceedings of ICWSM 2016 (2016).Giachanou, Anastasia, and Fabio Crestani.

[8] Jose Berengueres, Guillem Duran and Dani Castro. Happiness an inside Job? Turnover prediction from likeability, engament and relative happiness. IEEE/ACM ASONAM 2017 (in Press)

[9] Daniel Marco and David L. Neuhoff, "The Validity of the Additive Noise Model for Uniform Scalar Quantizers", IEEE Transactions on Information Theory, Vol. IT-51, No. 5, pp. 1739–1755, May 2005. doi:10.1109/TIT.2005.846397

[10] Boswell, W., Shipp, A., Payne, S. & Culbertson, S. (2009). Changes in newcomer job satisfaction over time: Examining the pattern of honeymoons and hangovers. Journal of Applied Psychology, 94 (4). 844-858.

[11] Thomson, David L., Rober W. Furness, and Pat Monaghan. "The analysis of ordinal response data in the behavioural sciences." Animal behaviour 56.4 (1998): 1041-1043